\documentclass[prl,twocolumn,showpacs]{revtex4}
\usepackage{array,amsmath,amssymb,graphicx,psfrag}
\newcount\minute
\newcount\hour
\newcount\hourMins
\def\now{
\minute=\time
\hour=\time \divide \hour by 60
\hourMins=\hour \multiply\hourMins by 60
\advance\minute by -\hourMins
\zeroPadTwo{\the\hour}:\zeroPadTwo{\the\minute}
}
\def\zeroPadTwo#1%
{%
\ifnum #1<10 0\fi
#1%
}

\begin{document}
\newcommand{\bnabla}{\bm{\nabla}}
\newcommand{\bz}{{\mathbf z}}
\newcommand{\bx}{{\mathbf x}}
\newcommand{\br}{{\mathbf r}}
\newcommand{\bG}{{\mathbf G}}
\newcommand{\bu}{{\mathbf u}}
\newcommand{\bq}{{\mathbf q}}
\newcommand{\cH}{{\cal H}}
\newcommand{\dif}{{\mathrm d}}

\title{\textsc{Transport properties of clean and disordered Josephson junction arrays}}

\author{Aleksandra Petkovi\' c$^{1,2}$, Valerii M. Vinokur$^{2}$, and Thomas Nattermann$^{1}$
}

\affiliation{$^1$ Institut f\"ur Theoretische Physik, Universit\"at zu
K\"oln, Z\"ulpicher Str. 77, 50937 K\"oln, Germany\\
$^2$Materials Science Division, Argonne National Laboratory,
Argonne, Illinois 60439 }

\date{\today,\now}

\begin{abstract}
We investigate the influence of quantum fluctuations and weak disorder on the
vortex dynamics in a two-dimensional superconducting
Berezinskii-Kosterlitz-Thouless system. The
temperature below which quantum fluctuations dominate the vortex
creep is determined, and the transport in this quantum regime is
described. The crossover from quantum to classical regime
  is discussed and the quantum correction to the classical current-voltage relation is
  determined. It is found that weak disorder can effectively reduce the critical current as compared
   to that in the clean system.
\end{abstract}

\maketitle

The physics of the Berezinskii-Kosterlitz-Thouless (BKT) transition in thin
superconducting films is re-emerging as one of the mainstreams in the
current condensed matter physics.
The interest is motivated by recent advances in studies of layered
high-temperature superconductors~\cite{Ruef-2006,Matthey-2007}, the discovery of the
superconductivity at the interface between the insulating oxides~\cite{Reyren-2007,Caviglia-2008},
new studies in thin superconducting films uncovering the role of
the two-dimensional (2D) superconducting fluctuations~\cite{Crane-2007,Pourret-2006},
and the intense developments in the physics
of the superconductor-insulator transition where the BKT transition may play a major role~\cite{Vinokur-Nature}.

The predicted benchmark of the transition that serves to detect
it experimentally is the change of the shape of the $I$-$V$
characteristics, $V\propto I^{1+\alpha}$ from $\alpha=0$ above the
transition, $T>T_{BKT}$, to $\alpha=2(T_{BKT}/T)$ at $T\leq
T_{BKT}$~\cite{HN,Minnhagen}. However, experimental data on
superconducting films show appreciable deviations from the
theoretical predictions and are still inconclusive~\cite{exp}.
Among the possible sources of the deviation from the classic
predictions, one can consider the finite size
effect,~\cite{Kogan,GV} and effects of
disorder~\cite{Na+_95,KoNa_96,Giam}. Another important issue is
the role of quantum effects which become crucial when the BKT
transition occurs at low enough temperatures. In this
Brief Report we will analyze the role of quantum effects in the
BKT transition paying a special attention to the intermediate region of
the interplay between thermal and quantum contributions.  We will
discuss the effect of disorder-generated vortices on the BKT
transition, neglecting quantum fluctuations, namely the effective
reduction of the critical current as compared to that in clean
samples.

\emph {Model.} We choose a disordered Josephson junction array as a convenient discrete model for the 2D disordered
superconducting film~\cite{EcSc_89}. The Hamiltonian describing the
system is:
   \begin{equation}\label{eq:H}
       {\cal H} = \frac{1}{2} \sum_{i,j} ( C^{-1})_{i,j}\hat n_i\hat n_j -J\sum_{\langle i,j\rangle}\cos (\hat \varphi_i-\hat \varphi_{j}-A_{ij})
   \end{equation}
where
$[\hat n_j, \hat \varphi_k]=-2e i\delta_{j,k}$.
We ignore single electron tunneling and other sources of dissipation.
The only non-vanishing elements of the capacitance matrix $C_{ij}$
are its diagonal elements $C_{jj} =4C$ (no summation over the
repeated index) and $C_{ij}=-C$ for the nearest neighbors $i,j$, i.e.,
the capacitance to the ground is assumed negligible as compared to the
mutual capacitances of the superconducting islands. The second sum
in (\ref{eq:H}) is taken over all nearest-neighbor pairs on a square
lattice. Random phase shifts $A_{ij}$ result from the
deviations of the flux in a distorted plaquette from an integer
multiple of the flux quantum $\Phi_0=\hbar c/2e$
\cite{Granato+86}.

In the clean classical case, i.e.~for $A_{ij}=0$ and in the limit $C\to \infty$, the physics of the system can be most adequately described in terms of vortices that experience the superconducting BKT transition at the temperature $T_{BKT}\simeq \pi \widetilde{J}/2$, where $\widetilde{J}$ denotes the renormalized coupling constant.
It is convenient to decompose the phase at the site $i$, as $\varphi_i = \varphi_i^{(v)}+\varphi_{i}^{(sw)}$ where $(v)$ and $(sw)$ stand for the vortex and the spin wave part, respectively.
Then, the vortex Hamiltonian can be written as
\begin{align}\label{eq:H_v}
\mathcal{H}_{v}=&-{J\pi} \sum_{i\neq j} m_{i}   m_j \ln{ \frac{| {\bf r}_{i}-{\bf r}_{j}|}{ \xi}} +\sum_i E_c m_i^2,
\end{align}
where $E_c$ denotes the core energy of a vortex. The sums are taken over the sites ${\bf r}_i$ of a dual lattice; $m_i$ is the vorticity of the $i$th vortex, and we assumed that $\sum_i m_i=0$, where $\xi$ denotes the superconductor coherence length.

Next we want to include quantum fluctuations. After going over to the path integral
representation of the partition function and integrating out the charge degrees of freedom, the action of the Josephson junction array in the limit $E_c=e^2/2C \ll J$ assumes the form~\cite{EcSc_89,fazio91}
\begin{align}
S=\int d\tau \left(\frac{M}{2}\sum_{i}(\partial_{\tau}\mathbf r_{i})^2
+{\cal H}_v\right).
\end{align}
The vectors ${\bf r}_i(\tau)$ are the world lines of the vortices and $M=h^2 C /(8 e^2 \xi^2)$.

\emph{Clean case}. We begin with the discussion of a clean case.
If we apply an external transport current, it will exert the force
${\bf f}\sim {\bf j}$ on the vortices, where $\bf{j}$ is the current
density \cite{Tinkham}. This generates an additional term $-\sum_i
m_i{\bf f}\cdot{\bf r}_i$ in (\ref{eq:H_v}). In order to describe
the effect of vortices on the current-voltage relation
quantitatively, we consider the effect of vortices crossing the
system transversely to the transport current. This motion
dissipates energy. The Bardeen-Stephen flux flow resistance \cite{Bardeen}
gives for the current-voltage ($V-I$) relation
\begin{equation}
V=2\pi\xi^2\rho_nn_vI
\end{equation}
where $n_v$ is the vortex density and $\rho_n$ is the normal state resistivity. The rate equation for the vortex density is
\begin{align}\label{eq:decayrate}
\partial_t n_v=\Gamma -\frac{\xi^2}{\tau_{rec}}n_v^2.
\end{align}
Here $\Gamma$ denotes the rate of generation of free vortices, while the second term on the rhs of (\ref{eq:decayrate}) describes their recombination, $\tau_{rec}$ denotes the recombination parameter. The steady state value $n_v=(\tau_{rec}\Gamma)^{1/2}/\xi$ of the vortex density determines the current-voltage relation.

In order to determine $\Gamma$, we consider the appearance of a
vortex-antivortex pair and its subsequent separation via tunneling
or thermal activation under the influence of the external force
$\mathbf{f}$. In the clean case this process is symmetric, i.e.,
the coordinates of the vortex $\br_1$ and the antivortex $\br_2$
satisfy ${\bf r}_1=-{\bf r}_2=\br$ with ${\mathbf {f}\cdot
\bf{r}}=f r$. The action of the vortex pair can be rewritten as
\begin{align}\label{ch2_eq:action}
&S=\int d\tau \left[M(\partial_{\tau}r)^2+U(r)\right ],
\end{align}
where $U(r)=2 \pi J \ln{\left( \frac{2r}{\xi} \right)}-2 f r+2 E_c$.
The problem effectively reduces to a single particle motion through one-dimensional potential barrier $U(r)$ \cite{footnote1}.

\begin{figure}[h]
\includegraphics[width=0.7\columnwidth]{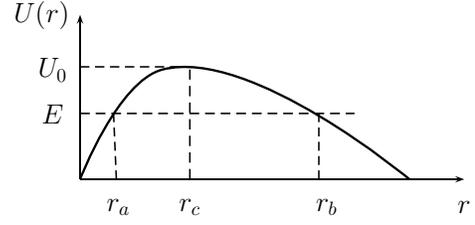}
\caption{Potential barrier for the separation of the vortex-antivortex pair.}\label{fig:potential}
\end{figure}

The rate $\Gamma$ is given by~\cite{Affleck81}
\begin{equation}\label{eq:Gamma-T}
\Gamma\sim \int_{0}^{\infty} dE \;\Gamma(E)e^{-E/T},
\end{equation}
where $\Gamma(E)$ denotes the zero temperature tunneling rate of a particle in the potential $U(r)$ having an energy $E$.
For low temperatures  and hence $E$ smaller than the barrier height $U_0=2\pi J\left[\ln(\frac{2\pi J}{f\xi})-1\right]+2E_c$, $\Gamma(E)$ in the WKB approximation is
\begin{align}
\Gamma(E)= e^{-4\sqrt{M}\int_{r_a(E)}^{r_b(E)}d r\sqrt{U(r)-E}/\hbar},
\end{align}
where $r_{a/b}(E)$ satisfy $U(r_{a/b})=E$ (see
Fig.~\ref{fig:potential}).
In the following different regimes will be considered.\\
(i)At zero temperature the only contribution in Eq.~(\ref{eq:Gamma-T}) comes from $E=0$. The generated voltage for small currents ($f\xi/J\ll 1$) is
\begin{align}\label{eq:quantum}
&V\sim\Gamma^{1/2}\sim e^{-S(0,0)/2\hbar}\notag\\
&\frac{S(0,0)}{\hbar}\approx c_1\frac{\sqrt{M}(2J\pi)^{3/2}}{\hbar f}{\left(\ln{\frac{2J\pi}{f\xi}}\right)}^{3/2}.
\end{align}
$c_1$ is a positive constant of the order of unity and
\begin{align}
\frac{S(E,T)}{\hbar}=\frac{E}{T}+4\sqrt{M}\int_{r_a(E)}^{r_b(E)}d r\frac{\sqrt{U(r)-E}}{\hbar}
\end{align}
is the action of the classical path of the particle in the potential $-U(r)$ with the energy $E$ and mass $2M$. The result  (\ref{eq:quantum}) is in an agreement with that of Ref.~\cite{Iengo+96} where it is obtained using the different technique~\cite{footnote2}.
We find that the result (\ref{eq:quantum}) holds also at finite temperatures as long as
\begin{align}\label{eq:To}
T\leq T_0=\frac{1}{c_2}\frac{\hbar f}{\sqrt{\pi 2J M}}\frac{1}{\sqrt{\ln{\frac{\pi 2J}{f\xi}}}},
\end{align}
where $c_2$ is positive constant of the order of unity.\\
(ii) At intermediate temperatures $T_0<T<T^*$, where
\begin{align}\label{eq:T^*}
T^*=\frac{\hbar}{2\pi}\sqrt{\frac{-U''(r_c)}{2M}}=\frac{\hbar
f}{2\pi}\sqrt{\frac{1}{M J \pi}},
\end{align}
the main contribution in Eq.~(\ref{eq:Gamma-T}) comes from the
stationary point $E_T$. Therefore,
$V\sim \exp \left [ -S(E_T,T)/2\hbar \right ]$.
$E_T$ depends on the temperature and is implicitly given by the
equation
\begin{align}\label{eq:finiteT}
\frac{\hbar}{T}=2\sqrt{M}\int_{r_a(E_T)}^{r_b(E_T)}d r
\frac{1}{\sqrt{U(r)-E_T}}=\tau(E_T),
\end{align}
where $\tau(E)$ can be interpreted as the period of the classical motion of a particle with the mass $2M$ and energy $E$, in the potential
$-U(r)$. Since $\tau(E)$ is the monotonically decreasing function of $E$ for small currents,
 Eq.~(\ref{eq:finiteT}) has the unique solution $E_T$ for every $T$ in a range $T_0\leq T\leq
 T^*$. We come back to the discussion of the voltage characteristic in
 this regime below.
\\(iii)At even higher temperatures $T^*< T \leq T_{\mathrm{BKT}}$, the decay rate is dominated by  $E>U_0$~\cite{Goldanskii,Affleck81} and the thermally
activated breaking of vortex pairs dominates the
dynamics. Then, the decay rate is given by the Arrhenius law $\Gamma\sim \exp[-S_{class}/ \hbar]$ where $S_{class}=\hbar U_0/T$. The voltage-current relation reads \cite{HN,Doniah+79}
\begin{align}\label{ch2_eq:VIclassical}
V\sim f e^{-U_0/(2T)} \sim j^{\delta(T)},\quad\quad \delta(T)=1+{\pi J}/{T}.
\end{align}
Taking into account the presence of other vortices by replacing $J\to \widetilde{J}$, the coefficient assumes a
universal value $\delta(T_{BKT})=3$.\\
(iv) At $T>T_{\mathrm{BKT}}$ a finite density of free vortices appears in an equilibrium, and the system is characterized by a linear current-voltage relation for small enough currents.

\begin{figure}\center
\includegraphics[width=0.8\linewidth]{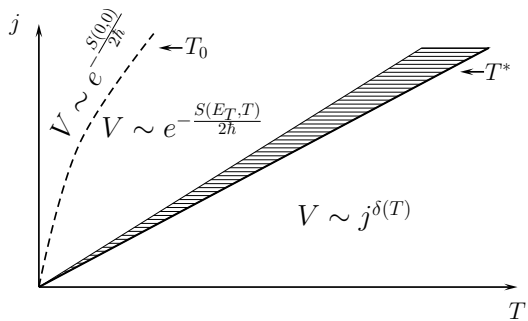}
\caption{Dynamic phase diagram in current-temperature coordinates
showing different types $V(j,T)$ dependencies for $T<T_{BKT}$. The
dashed and the solid lines sketch $T_0(j)$ and $T^*(j)$,
respectively.  In the domain $T<T_0$ quantum tunneling of vortices
dominates the vortex dynamics, while at $T>T^*$ the
voltage-current characteristics is determined by the thermally
activated motion. In the shaded region the quantum correction
 to the classical result, given by Eq.~(\ref{eq:correction}), applies.}\label{Fig:regions}
\end{figure}
Next, we consider crossover from the quantum- ($T\leq T_0$) to the classical regime ($T>T^*$) in more detail. Within the semiclassical approximation the decay rate is given, with the exponential accuracy, by $\Gamma\sim \exp[-S_{\mathrm{min}}/\hbar]$, where $S_{\mathrm{min}}$ is the action of the trajectory minimizing the Euclidean action of Eq.~(\ref{ch2_eq:action}). For temperatures below $T_0$ the extremal action is $S_{\mathrm{min}}=S(0,0)$, in the intermediate region ($T_0<T<T^*$) the minimal action is $S_{\mathrm{min}}=S(E_T,T)$, and in the high temperature regime the trajectory extremizing the action is time independent, and therefore $S_{\mathrm{min}}=\hbar U_0/T$.
We find that $S_{\mathrm{min}}$ at $T^*$ has a continuous first derivative with respect to temperature, while the second derivative has a jump:
\begin{align}\label{ch2_eq:secondorder}
&\frac{\dif S(E_T,T)}{\dif T}\Big|_{T^*}=\frac{\dif S_{\mathrm{class}}}{\dif T}\Big|_{T^*}\notag\\ &\frac{\dif^2 S(E_T,T)}{\dif T^2}\Big|_{T^*} < \frac{\dif^2 S_{\mathrm{class}}}{\dif T^2}\Big|_{T^*}.
\end{align}
Following Ref.~\cite{Larkin+83} we call this situation a
second-order transition at the crossover point \cite{footnote3}.
The result of Eqs.~(\ref{ch2_eq:secondorder}) is a general
property of a massive particle trapped in a metastable state
formed by a potential $U(r)$, provided $\tau(E)$ is a monotonously
decreasing function of energy \cite{Chudnovsky92}.

Generally, in the case of a second-order transition the trajectory extremizing the action can be written as~\cite{Larkin+83}
\begin{align}
r(\tau)=r_c+\sum_{n=1}^{\infty}a_n \cos{\left( \frac{2\pi
T}{\hbar}n \tau\right)},
\end{align}
where the coefficients $|a_n|\ll |a_1|$ ($n>1$) are small near the
transition temperature $T^*$. Substituting $r(\tau)$ in
Eq.~(\ref{ch2_eq:action}), the action can be expanded in powers of
$a_n$, yielding an effective action $
S\approx {U_0\hbar}/{T}+\alpha a_1^2+\beta a_1^4,
$
where the coefficient $\alpha$ is negative below $T^*$ and vanishes at the transition temperature $T^*$ \cite{Larkin+83}.
Then the coefficient $a_1$ can be found from the minimization of the action $S$ and the minimal action is
$
S_{\mathrm{min}}=U_0\hbar/T-\alpha^2/(4\beta).
$
Following Refs.~\cite{Larkin+83}, we determine the coefficients $\alpha$ and $\beta$ and find a quantum correction to the classical result of Eq.~(\ref{ch2_eq:VIclassical})
\begin{align}\label{eq:correction}
V&\sim j^{\delta(T)} e^{\Delta},\notag\\
\Delta & =
\frac{(T^2-T^{*2})^2}{T T^{*^3}} \frac{\sqrt{M J^3}  }{\hbar f} \frac{\pi^{5/2}}{1+2(1-4(T/T^*)^2)^{-1}} .
\end{align}
This result is valid near the transition, for temperatures approaching $T^*$ from below, see Fig.~\ref{Fig:regions}. We conclude that quantum effects significantly enhance the decay rate in comparison to the classical rate for the asymptotically small currents.
It would be interesting to probe the result of Eq.~(\ref{eq:correction}) in
experiments.
\\
\emph{Disordered case}. Next we include disorder into the
consideration, in the limit $C\to \infty$. The phase shifts are
assumed to be uncorrelated from bond to bond, and each is Gaussian
distributed with the mean value and the variance
\begin{equation}\label{eq:A}
\overline {A_{ij}} = 0,\quad\quad\overline {A_{ij}^2} =\sigma,
\end{equation}
respectively. Then, an additional term $\sum _i m_iV({\bf r}_i)$ is generated in (\ref{eq:H_v}), where $V({\bf r}_i)=2\pi J\sum_j Q_j\ln(|
{\bf r}_{i}-{\bf r}_{j}| / \xi)$. $Q_i=(1/2\pi)\sum_{<plaq>} A_{ij}$
are the frozen charges sitting on the dual lattice in the center of
a plaquette  whereas the sum is over a plaquette formed by 4 bonds. From (\ref{eq:A}) follows $\overline{(V({\bf r}_i)-V({\bf r}_j))^2}=4\pi\sigma J^2\ln (| {\bf r}_{i}-{\bf r}_{j}|/ \xi)$.

It was shown in Ref.~\cite{Na+_95}, that the system in the classical case, at $T = 0$ undergoes a disorder driven transition from the `ordered' BKT
state to a disordered phase at the critical disorder strength
$\sigma_c=\pi/8$. In the ordered BKT phase vortices appear,
on average, only in a form of the bound pairs. Indeed, the energy of a vortex pair
with the separation $R$ and $m_1=-m_2=1$ in a clean sample is given by
$2 \pi J \ln (R/ \xi)$. Since $V({\bf r}_i)$ is Gaussian distributed, the
typical energy gain is $-2J \sqrt {\pi \sigma \ln(R/ \xi)}$ which is
smaller by a factor $\sim (\ln (R/\xi))^{-1/2}$ than the energy cost of a
pair. However, the maximum energy gain of a vortex dipole in a
region of linear size $L>R$ is larger by a factor $\sqrt{2\ln N}$ than the typical energy gain,
which arises from the $N$ independent realizations of the
vortex positions \cite{KoNa_96}. The disorder potential, that one vortex-antivortex pair of size $R+dR$ feels, is uncorrelated when the pair is translated over a distance larger than $R$  \cite{LHT96}. Therefore, we introduce a lattice with a lattice constant $R$. Since also the correlations of disorder potential inside the cell give only subleading-order corrections \cite{LHT96}, we estimate $N\approx (L/R)^2 (R/\xi)^2 (2\pi R/\xi) dR/\xi$. For $dR\approx R$, we get the free energy of the pair at $T=0$
\begin{align}\label{eq:pairenergy}
E\approx 2\pi J \ln{\frac{R}{\xi}}\left[1-\sqrt{\frac{4\sigma}{\pi} \frac{\ln{(L R/ \xi^2)}}{\ln{(R/\xi)}}}\right].
\end{align}
Thus, if $R\approx L$,  the total energy of the coresponding vortex pair becomes
negative and free vortices are favored by disorder provided $\sigma
>\sigma_c=\pi/8$, in an agreement with the renormalization group result in Ref.~\cite{Na+_95}.
Note that strictly speaking these vortices are ``pseudo-free" since
despite the fact that their attraction is overruled by disorder,
they remain pinned by the same disorder-induced forces.
It follows from the above reasoning that even for $\sigma<\sigma_c$
some rare vortex pairs of the negative energy can appear. From (\ref{eq:pairenergy}) we get that their maximal size is $R_c\approx \xi (L/\xi)^{\frac{1}{2\sigma_c/\sigma-1}}$, which reaches the size of the system for
$\sigma\to \sigma_c-0$, as expected. Typically there is a single
dipole of the size $R_c$ in the system. If we divide the system into
$M^2$ subsystems, each part will contain a dipole of the maximum  size $R_{M}
\approx R_c M^{-{\frac{1}{2\sigma_c/\sigma-1}}}$. The density of dipoles of the size
$R_M$ is $\xi^{-2}(R_M/ \xi)^{2(1-2\sigma_c/\sigma)}$ at $T=0$, in agreement with  Ref.~\cite{LHT96}.

We further determine the critical current. If the transport current is strong enough, it will
depin vortices such that the dissipation sets in. A crude estimate
for the critical depinning force at $T=0$ and $\sigma<\sigma_c$ is given by
\begin{equation}\label{eq:criticalforce}
f_c\sim \frac{J}{R_c} \sim \frac{J}{\xi}\left(\frac{L}{\xi}\right)^{\frac{-1}{2\sigma_c/\sigma-1}},
\end{equation}
since smaller dipoles are depinned at larger forces. The influence of disorder on the voltage-current relation is left for further studies.

\emph{Conclusion.} We have investigated transport properties of Josephson
junction arrays taking into account the influence of quantum
fluctuations on the unbinding of vortex pairs for $E_c\ll J$. At
sufficiently low temperatures the quantum tunneling of vortices
turns out to be more probable than their thermal activation. We have
derived the $V$-$I$ relation corresponding to the quantum creep of the BKT-vortices and found the range
of temperatures, $0\leq T\leq T_0$, where this law is applicable. We
have determined the temperature $T^*$ above which the thermally
activated breaking of vortex pairs dominates the vortex
nucleation. We have discussed the region of intermediate
temperatures $T_0<T<T^*$ where a crossover from classical to
quantum behavior occurs, and found the quantum correction to the
classical result, see Eq.~(\ref{eq:correction}). The results are
schematically summarized in Fig.~\ref{Fig:regions} and can be straightforwardly extended to the quantum limit
$E_c\gg J$, where the transport is mediated by the motion of charges
dual to the superconducting vortices, via the standard dual
transformation. Moreover, in the presence of positional disorder
and for $C\to \infty$, we have shown that additional vortices
generated by the disorder contribute to transport, effectively
reducing the critical current as compared to that in a clean
system.

We are delighted to thank R. Fazio and Z. Ristivojevic for useful discussion. This
work was supported by the U.S. Department of Energy Office of
Science through contract No. DE-AC02-06CH11357; authors like to
acknowledge the support from the SFB 608 (AP and TN) and the AvH
foundation (VMV).



\begin{thebibliography}{99}
\bibitem{Ruef-2006} A.\,R\"ufenacht, \textit{et al}, Phys. Rev. Lett., \textbf{96} 227002 (2006).
\bibitem{Matthey-2007} D.\,Matthey, N.\,Reyren, J.-M.\,Triscone, and T.\,Schneider,
Phys. Rev. Lett., \textbf{98}, 057002 (2007).
\bibitem{Reyren-2007} N.\,Reyren \textit{et al}, Science, \textbf{317}, 1196 (2007).
\bibitem{Caviglia-2008} A.\,D.\,Caviglia, \textit{et al}, Nature \textbf{456}, 624 (2008).
\bibitem{Crane-2007} R.\,W.\,Crane, \textit{et al}, Phys. Rev \textbf{B 75}, 094506 (2007).
\bibitem{Pourret-2006} A.\,Pourret, \textit{et al}, Nat. Phys. \textbf{2}, 683 (2006).
\bibitem{Vinokur-Nature} V. M. Vinokur \textit{et al}, Nature (London) \textbf{452}, 613 (2008).


\bibitem{HN}B.\,I.\,Halperin and D.\,R.\,Nelson, J. Low Temp. Phys. \textbf{36},
599 (1979).

\bibitem{Minnhagen} P.\,Minnhagen, Rev. Mod. Phys. \textbf{59}, 1001 (1987).

\bibitem{exp} J.\,M.\,Repaci et al., Phys. Rev. B \textbf{54}, R9674 (1996); D. R. Strachan,
C.\,J.\,Lobb, and R.\,S.\,Newrock, Phys. Rev. B \textbf{67}, 174517
(2003); M.\,M.\,Ozer, J.\,R.\,Thompson, and H.\,H.\,Weitering, Phys. Rev. B \textbf{74}, 235427 (2006);
M.\,M.\,Ozer et al., Science \textbf{316}, 1594 (2007); A.
R\"{u}fenacht et al., Phys. Rev. Lett. \textbf{96}, 227002 (2006);
F.\,Tafuri et al., Europhys. Lett. \textbf{73}, 948 (2006).

\bibitem{Kogan} V. G. Kogan, Phys. Rev. B \textbf{75}, 064514 (2007).

\bibitem{GV} A.\,Gurevich and V.\,M.\,Vinokur, Phys. Rev. Lett. \textbf{100}, 227007 (2008).
\bibitem{Na+_95} T.\,Nattermann, S.\,Scheidl, S.\,E.\,Korshunov and M.\,S.\,Li, J. Phys. (France) I, {\bf 5}, 565 (1995).

\bibitem{KoNa_96} S.\,E.\,Korshunov and T.\,Nattermann, Phys. Rev. B {\bf  53}, 2746 (1996), Physica B {\bf 222}, 280 (1996).
\bibitem{Giam} L.\,Benfatto, C.\,Castellani and T.\,Giamarchi, arxiv:0909.0479

\bibitem{EcSc_89} U.\,Eckern and A.\,Schmid, Phys. Rev. B {\bf 39}, 6441 (1989).

\bibitem{Granato+86} E.\,Granato and J.\,M.\,Kosterlitz, Phys. Rev. B {\bf 33}, 6533 (1986).

\bibitem{fazio91} R.\,Fazio and G.\,Sch\" on, Phys. Rev. \textbf{B 43}, 5307 (1991).

\bibitem{Tinkham} M.\,Tinkham, ``\emph{Introduction to
Superconductivity}'' Dover publications
2004.

\bibitem{Bardeen} J.\,Bardeen and  M.\,J.\,Stephen, Phys. Rev. {\bf 140} 1197A (1965).

\bibitem{footnote1}{Note that the fact that the system is two-dimensional is important for pre-exponential factors in the vortex production rate $\Gamma$. However, in the following we will determine $\Gamma$ with an exponential accuracy since we are interested in asymptotically small currents.}

\bibitem{Affleck81} I.\,Affleck, Phys. Rev. Lett. \textbf{46}, 388 (1981).

\bibitem{Iengo+96} R.\,Iengo and G.\,Jug, Phys. Rev. B {\bf 54}, 13207 (1996).

 \bibitem{footnote2}{In P.\,Ao, J. Low Temp. Phys. {\bf 89}, 543 (1992), the same leading dependence on $f$ is found, but a different logarithmic dependence on $f$.}

\bibitem{Goldanskii} V.\,I.\,Goldanskii, Dokl. Acad. Nauk SSSR \textbf{124}, 1261 (1959) [Sov. Phys. Dokl. \textbf{4},74 (1959)].


\bibitem{Doniah+79} S.\,Doniach and B.\,A.\,Huberman, Phys. Rev. Lett. {\bf 42}, 1169 (1979).

\bibitem{Larkin+83} A.\,I.\,Larkin and Yu.\,N.\,Ovchinnikov, JETP Lett. {\bf 37}, 382 (1983), Sov. Phys. JETP {\bf 59}, 2 (1984).

\bibitem{footnote3}{Note that quantum fluctuations smear the transition in a narrow temperature region near $T^*$ \cite{Chudnovsky92}.}

\bibitem{Chudnovsky92} E.\,M.\,Chudnovsky, Phys. Rev. A {\bf 46}, 8011 (1992).



\bibitem{LHT96} L.-H.\,Tang, Phys. Rev. B {\bf 54}, 3350 (1996).

\end{thebibliography}
\end{document}